\documentclass[aps,pre,showpacs,groupedaddress]{revtex4}
\usepackage{amsfonts}
\usepackage{mathrsfs}
\usepackage{graphicx}
\usepackage{amsmath}
\usepackage{amssymb}
\usepackage{amsbsy}

\begin{document}

\title{Sparse Hopfield network reconstruction with $\ell_{1}$ regularization}

\author{Haiping Huang}
\email{physhuang@gmail.com}
\affiliation{Department of Computational
Intelligence and Systems Science, Tokyo Institute of Technology,
Yokohama 226-8502, Japan}
\date{\today}

\begin{abstract}
We propose an efficient strategy to infer sparse Hopfield network
based on magnetizations and pairwise correlations measured through
Glauber samplings. This strategy incorporates the $\ell_{1}$
regularization into the Bethe approximation by a quadratic
approximation to the log-likelihood, and is able to further reduce
the inference error of the Bethe approximation without the
regularization. The optimal regularization parameter is observed to
be of the order of $M^{-\nu}$ where $M$ is the number of independent
samples. The value of the scaling exponent depends on the
performance measure. $\nu\simeq0.5001$ for root mean squared error
measure while $\nu\simeq0.2743$ for misclassification rate measure.
The efficiency of this strategy is demonstrated for the sparse
Hopfield model, but the method is generally applicable to other
diluted mean field models. In particular, it is simple in
implementation without heavy computational cost.

\end{abstract}

 \maketitle

\section{Introduction}
The inverse Ising problem is intensively studied in statistical
physics, computational biology and computer science in the few past
years~\cite{Nature-06,Cocco-09,Weigt-11,Wain-2010}. The biological
experiments or numerical simulations usually generate a large amount
of experimental data, e.g., $M$ independent samples
$\{\boldsymbol{\sigma}^{1},\boldsymbol{\sigma}^{2},\ldots,\boldsymbol{\sigma}^{M}\}$
in which $\boldsymbol{\sigma}$ is an $N$-dimensional vector with
binary components ($\sigma_{i}=\pm1$) and $N$ is the system size.
The least structured model to match the statistics of the
experimental data is the Ising model~\cite{Bialek-09ep}:
\begin{equation}\label{Ising}
    P_{{\rm
Ising}}(\boldsymbol{\sigma})=\frac{1}{Z(\mathbf{h},\mathbf{J})}\exp\left[\sum_{i<j}J_{ij}\sigma_{i}\sigma_{j}+\sum_{i}h_{i}\sigma_{i}\right]
\end{equation}
where the partition function $Z(\mathbf{h},\mathbf{J})$ depends on
the $N$-dimensional fields and $\frac{N(N-1)}{2}$-dimensional
couplings. These fields and couplings are chosen to yield the same
first and second moments (magnetizations and pairwise correlations
respectively) as those obtained from the experimental data. The
inverse temperature $\beta=1/T$ has been absorbed into the strength
of fields and couplings.

Previous studies of the inverse Ising problem on Hopfield
model~\cite{Huang-2010a,Huang-2010b,SM-11,Pan-11,Huang-2012} lack a
systematic analysis for treating sparse networks. Inference of the
sparse network also have important and wide applications in modeling
vast amounts of biological data. Actually, the real biological
network is not densely connected. To reconstruct the sparse network
from the experimental data, an additional penalty term is necessary
to be added into the cost function, as studied in recovering sparse
signals in the context of compressed
sensing~\cite{Kabashima-2009,Montanari-2010,Elad-2010} or in Ising
model selection~\cite{Wain-2010,Bento-2011}. This strategy is known
as $\ell_{1}$-regularization which introduces an $\ell_{1}$-norm
penalty to the cost function (e.g., the log-likelihood of the Ising
model). The regularization is able to minimize the impact of finite
sampling noise, thus avoid the overfitting of data. The
$\ell_{1}$-regularization has been studied in the pseudo-likelihood
approximation to the network inference problem\cite{Aurell-2011} and
in the setting of sparse continuous perceptron memorization and
generalization~\cite{Weigt-09}. This technique has also been
thoroughly discussed in real neural data analysis using selective
cluster expansion method~\cite{Cocco-12,CM-12}. The cluster
expansion method involves repeated solution of the inverse Ising
problem and the computation of the cluster entropy included in the
expansion (cluster means a small subset of spins). To truncate the
expansion, clusters with small entropy in absolute value are
discarded and the optimal threshold needs to be determined.
Additionally, the cluster size should be small to reduce the
computational cost while at each step a convex optimization of the
cost function (see Eq.~(\ref{cost})) for the cluster should be
solved. This may be complicated in some cases. The pseudo-likelihood
maximization~\cite{Aurell-2011} method relies on the complete
knowledge of the sampled configurations, and involves a careful
design of the numerical minimization procedure for the
pseudo-likelihood (e.g., Newton descent method, or interior point
method) at a large computational cost (especially for large sample
size). In this paper, we provide an alternative way to reconstruct
the sparse network by combining the Bethe approximation and the
$\ell_{1}$-regularization, which is much simpler in practical
implementation. We expect that the $\ell_{1}$-regularization will
improve the prediction of the Bethe approximation. To show the
efficiency, we apply this method to the sparse Hopfield network
reconstruction.

Our contributions in this work are two-fold. (1) We provide a
regularized quadratic approximation to the negative log-likelihood
function for the sparse network construction by neglecting higher
order correlations, which yields a new inference equation reducing
further the inference error. Furthermore, the implementation is much
simple by saving the computational time. (2) Another significant
contribution is a scaling form for the optimal regularization
parameter is found, and this scaling form is useful for choosing the
suitable regularization. Most importantly, the method is not limited
to the tested model (sparse Hopfield model), and is generally
applicable to other diluted mean field models and even real data
analysis (e.g., neural data). The outline of the paper is as
follows. The sparse Hopfield network is defined in
Sec.~\ref{sec_sHopf}. In Sec.~\ref{sec_method}, we present the
hybrid inference method by using the Bethe approximation and
$\ell_{1}$-regularization. We test our algorithm on single instances
in Sec.~\ref{sec_result}. Concluding remarks are given in
Sec.~\ref{sec_Sum}.

\section{Sparse Hopfield model}
\label{sec_sHopf}

The Hopfield network has been proposed in Ref.~\cite{Hopfield-1982}
as an abstraction of biological memory storage and was found to be
able to store an extensive number of random unbiased
patterns~\cite{Amit-1987}. If the stored patterns are dynamically
stable, then the network is able to provide associative memory and
its equilibrium behavior is described by the following Hamiltonian:
\begin{equation}\label{Hami}
    \mathcal{H}=-\sum_{i<j}J_{ij}\sigma_{i}\sigma_{j}
\end{equation}
where the Ising variable $\sigma$ indicates the active state of the
neuron ($\sigma_{i}=+1$) or the silent state ($\sigma_{i}=-1$). For
the sparse network storing $P$ random unbiased binary patterns, the
symmetric coupling is constructed~\cite{Somp-1986,Fukai-1998} as
\begin{equation}\label{J_spar}
    J_{ij}=\frac{l_{ij}}{l}\sum_{\mu=1}^{P}\xi_{i}^{\mu}\xi_{j}^{\mu}
\end{equation}
where $l$ is the average connectivity of the neuron. $l\sim\mathcal
{O}(1)$ independent of the network size $N$. Note that in this case,
the number of stored patterns can only be finite. In the
thermodynamic limit, $P$ scales as $P=\alpha l$ where $\alpha$ is
the memory load. No self-interactions are assumed and the
connectivity $l_{ij}$ obeys the distribution:
\begin{equation}\label{distri}
    P(l_{ij})=\left(1-\frac{l}{N-1}\right)\delta(l_{ij})+\frac{l}{N-1}\delta(l_{ij}-1).
\end{equation}

Mean field properties of the sparse Hopfield network have been
discussed within replica symmetric approximation in
Refs.~\cite{Coolen-2003,Skantzos-2004}. Three phases (paramagnetic,
retrieval and spin glass phases) have been observed in this sparsely
connected Hopfield network with arbitrary finite $l$. For large $l$
(e.g., $l=10$), the phase diagram resembles closely that of
extremely diluted
($\lim_{N\rightarrow\infty}l^{-1}=\lim_{N\rightarrow\infty}l/N=0$,
such as $l=\ln N$) case~\cite{Watkin-1991,Canning-1992} where the
transition line between paramagnetic and retrieval phase is $T=1$
for $\alpha\leq 1$ and that between paramagnetic and spin glass
phase $T=\sqrt{\alpha}$ for $\alpha\geq 1$. The spin glass/retrieval
transition occurs at $\alpha=1$.

To sample the state of the original model Eq.~(\ref{Hami}), we apply
the Glauber dynamics rule:
\begin{equation}\label{GDrule}
    P(\sigma_{i}\rightarrow-\sigma_{i})=\frac{1}{2}\left[1-\sigma_{i}\tanh\beta h_{i}\right]
\end{equation}
where $h_{i}=\sum_{j\neq i}J_{ij}\sigma_{j}$ is the local field
neuron $i$ feels. In practice, we first randomly generate a
configuration which is then updated by the local dynamics rule
Eq.~(\ref{GDrule}) in a randomly asynchronous fashion. In this
setting, we define a Glauber dynamics step as $N$ proposed flips.
The Glauber dynamics is run totally $3\times 10^{6}$ steps, among
which the first $1\times 10^{6}$ steps are run for thermal
equilibration and the other $2\times 10^{6}$ steps for computing
magnetizations and correlations, i.e.,
${m_{i}=\left<\sigma_{i}\right>_{{\rm data}},
C_{ij}=\left<\sigma_{i}\sigma_{j}\right>_{{\rm data}}-m_{i}m_{j}}$
where $\left<\cdots\right>_{{\rm data}}$ denotes the average over
the collected data. The state of the network is sampled every $20$
steps after thermal equilibration (doubled sampling frequency yields
the similar inference result), which produces totally $M=100 000$
independent samples. The magnetizations and correlations serve as
inputs to our following hybrid inference algorithm.

\section{Bethe approximation with $\ell_{1}$ regularization}
\label{sec_method}

The Bethe approximation assumes that the joint probability
(Boltzmann distribution, see Eq.~(\ref{Ising})) of the neuron
activity can be written in terms of single-neuron marginal for each
single neuron and two-neuron marginal for each pair of adjacent
neurons as
\begin{equation}\label{Bethe}
    P_{{\rm
Ising}}(\boldsymbol\sigma)\simeq\prod_{(ij)}\frac{P_{ij}(\sigma_{i},\sigma_{j})}{P_{i}(\sigma_{i})P_{j}(\sigma_{j})}\prod_{i}P_{i}(\sigma_{i})
\end{equation}
where $(ij)$ runs over all distinct pairs of neurons. This
approximation is exact on tree graphs and asymptotically correct for
sparse networks or networks with sufficiently weak
interactions~\cite{Mezard-09}. Under this approximation, the free
energy ($-\ln Z$) can be expressed as a function of connected
correlations $\{C_{ij}\}$ (between neighboring neurons) and
magnetizations $\{m_{i}\}$. The stationary point of the free energy
with respect to the magnetizations yields the following
self-consistent equations:
\begin{equation}\label{m}
m_{i}=\tanh\left(h_{i}+\sum_{j\in\partial
i}\tanh^{-1}\left(t_{ij}f(m_{j},m_{i},t_{ij})\right)\right)
\end{equation}
where $\partial i$ denotes neighbors of $i$, $t_{ij}=\tanh J_{ij}$
and
$f(x,y,t)=\frac{1-t^{2}-\sqrt{(1-t^{2})^{2}-4t(x-yt)(y-xt)}}{2t(y-xt)}$.
Using the linear response relation to calculate the connected
correlations for any pairs of neurons, we obtain the Bethe
approximation (BA) to the inverse Ising
problem~\cite{Ricci-2012,Berg-2012}:
\begin{equation}\label{BA}
  J_{ij}=-\tanh^{-1}\Biggl[\frac{1}{2(\mathbf{C}^{-1})_{ij}}(a_{ij}-b_{ij})
   -m_{i}m_{j}\Biggr],
\end{equation}
where $\mathbf{C}^{-1}$ is the inverse of the connected correlation
matrix, $a_{ij}=\sqrt{1+4L_{i}L_{j}(\mathbf{C}^{-1})_{ij}^{2}}$,
$L_{i}=1-m_{i}^{2}$ and
$b_{ij}=\sqrt{\left(a_{ij}-2m_{i}m_{j}(\mathbf{C}^{-1})_{ij}\right)^{2}-4(\mathbf{C}^{-1})_{ij}^{2}}$.
The couplings have been scaled by the inverse temperature $\beta$.
Note that fields can be predicted using Eq.~(\ref{m}) after we get
the set of couplings. Hereafter we consider only the reconstruction
of the coupling vector. In fact, the BA solution of the couplings
corresponds to the fixed point of the susceptibility
propagation~\cite{Mezard-09,Huang-2010b}, yet it avoids the
iteration steps in susceptibility propagation and the possible
non-convergence of the iterations. It was also found that the BA
yields a good estimate to the underlying couplings of the Hopfield
network~\cite{Huang-2010b}. In the following analysis, we try to
improve the prediction of BA with $\ell_{1}$-regularization.

The cost function to be minimized in the inverse Ising problem can
be written as the following rescaled negative log-likelihood
function~\cite{SM-09}:
\begin{equation}\label{cost}
\begin{split}
    S(\mathbf{h},\mathbf{J}|\mathbf{m},\mathbf{C})&=-\frac{1}{M}\ln\left[\prod_{\mu=1}^{M}P_{{\rm Ising}}(\boldsymbol{\sigma}^{\mu}|\mathbf{h},\mathbf{J})\right]\\
    &=\ln
    Z(\mathbf{h},\mathbf{J})-\mathbf{h}^{T}\mathbf{m}-\frac{1}{2}{\rm tr}(\mathbf{J}\mathbf{\tilde{C}})
    \end{split}
\end{equation}
where $m_{i}=\left<\sigma_{i}\right>_{{\rm data}}$ and $
\tilde{C}_{ij}=\left<\sigma_{i}\sigma_{j}\right>_{{\rm data}}$.
$\mathbf{h}^{T}$ denotes the transpose of the field vector while
${\rm tr}(\mathbf{A})$ denotes the trace of matrix $\mathbf{A}$. The
minimization of $S(\mathbf{h},\mathbf{J}|\mathbf{m},\mathbf{C})$ in
the $\frac{N(N+1)}{2}$-dimensional space of fields and couplings
yields the following equations:
\begin{subequations}\label{mJ}
\begin{align}
m_{i}&=\left<\sigma_{i}\right>,\\
C_{ij}&=\left<\sigma_{i}\sigma_{j}\right>-\left<\sigma_{i}\right>\left<\sigma_{j}\right>\label{mJ2}
\end{align}
\end{subequations}
where the average is taken with respect to the Boltzmann
distribution Eq.~(\ref{Ising}) with the optimal fields and couplings
(corresponding to the minimum of $S$). Actually, one can use Bethe
approximation to compute the connected correlation in the right-hand
side of Eq.~(\ref{mJ2}), which leads to the result of
Eq.~(\ref{BA}).

To proceed, we expand the cost function around its minimum with
respect to the fluctuation of the coupling vector up to the second
order as
\begin{equation}\label{expand}
 S(\mathbf{J})\simeq
 S(\mathbf{J}_{0})+\nabla S(\mathbf{J}_{0})^{T}\mathbf{\tilde{J}}+\frac{1}{2}\mathbf{\tilde{J}}^{T}\mathbf{H}_{S}(\mathbf{J}_{0})\mathbf{\tilde{J}}
\end{equation}
where $\mathbf{\tilde{J}}$ defines the fluctuation
$\mathbf{\tilde{J}}\equiv\mathbf{J}-\mathbf{J}_{0}$ where
$\mathbf{J}_{0}$ is the (near) optimal coupling vector. $\nabla
S(\mathbf{J}_{0})$ is the gradient of $S$ evaluated at
$\mathbf{J}_{0}$, and $\mathbf{H}_{S}(\mathbf{J}_{0})$ is the
Hessian matrix. The quadratic approximation to the log-likelihood
has also been used to develop fast algorithms for estimation of
generalized linear models with convex
penalties~\cite{Friedman-2010}. We have only made explicit the
dependence of $S$ on the coupling vector. The first order
coefficient vanishes due to Eq.~(\ref{mJ}). Note that the Hessian
matrix is an $N(N-1)/2\times N(N-1)/2$ symmetric matrix whose
dimension is much higher than that of the connected correlation
matrix. However, to construct the couplings around neuron $i$, we
consider only the neuron $i$-dependent part, i.e., we set $l=i$ in
the Hessian matrix
$\chi_{ij,kl}=\left<\sigma_{i}\sigma_{j}\sigma_{k}\sigma_{l}\right>-\left<\sigma_{i}\sigma_{j}\right>\left<\sigma_{k}\sigma_{l}\right>$
where $ij$ and $kl$ run over distinct pairs of neurons. This
simplification reduces the computation cost but still keeps the
significant contribution as proved later in our simulations. Finally
we obtain
\begin{equation}\label{apprS}
 S(\mathbf{J})\simeq
 S(\mathbf{J}_{0})+\frac{1}{2}\sum_{ij,ki}\tilde{J}_{ij}(\tilde{C}_{jk}-\tilde{C}_{ij}\tilde{C}_{ki})\tilde{J}_{ki}+\lambda\sum_{ij}|J_{0,ij}+\tilde{J}_{ij}|
\end{equation}
where an $\ell_{1}$-norm penalty has been added to promote the
selection of sparse network
structure~\cite{Hertz-12,Cocco-12,Bento-2011}. $\lambda$ is a
positive regularization parameter to be optimized to make the
inference error (see Eq.~(\ref{error})) as low as possible. The
$\ell_{1}$-norm penalizes small but non-zero couplings and
increasing the value of the regularization parameter $\lambda$ makes
the inferred network sparser. In the following analysis, we assume
$\mathbf{J}_{0}$ is provided by the BA solution (a good
approximation to reconstruct the sparse Hopfield
network~\cite{Huang-2010b}, yielding a low inference error), then we
search for the new solution to minimize the regularized cost
function Eq.~(\ref{apprS}), finally we get the new solution as
follows,
\begin{equation}\label{reg}
 J^{(i)}_{ij}=J_{0,ij}-\lambda\sum_{k}{\rm sgn}(J_{0,ik})[\mathbf{C}^{i}]^{-1}_{kj}
\end{equation}
where ${\rm sgn}(x)=x/|x|$ for $x\neq0$ and
$(\mathbf{C}^{i})_{kj}=\tilde{C}_{kj}-\tilde{C}_{ji}\tilde{C}_{ik}$.
Eq.~(\ref{reg}) results from $\frac{\partial S(\mathbf{J})}{\partial
J_{ij}}=0$ which gives
$\mathbf{\tilde{J}}^{T}\mathbf{C}^{i}=\boldsymbol{\Lambda}^{T}$,
where $\Lambda_{j}=-\lambda{\rm sgn}(J_{0,ij}) (j\neq i)$ and
$\tilde{J}_{j}=J_{ij}-J_{0,ij} (j\neq i)$. $J^{(i)}_{ij}$ represents
couplings around neuron $i$. To ensure the symmetry of the
couplings, we construct
$J_{ij}=\frac{1}{2}(J^{(i)}_{ij}+J^{(j)}_{ji})$ where $J^{(j)}_{ji}$
is also given by Eq.~(\ref{reg}) in which $i$ and $j$ are exchanged.
The inverse of $\mathbf{C}^{i}$ or $\mathbf{C}^{j}$ takes the
computation time of the order $\mathcal {O}(N^{3})$, much smaller
than that of the inverse of a susceptibility matrix
$\boldsymbol{\chi}$.

We remark here that minimizing the regularized cost function
Eq.~(\ref{apprS}) corresponds to finding the optimal deviation
$\mathbf{\tilde{J}}$ which provides a solution to the regularized
cost function. We also assume that for small $\lambda$, the
deviation is small as well. Without the quadratic approximation in
Eq.~(\ref{expand}), no closed form solution exists for the optimal
$\mathbf{J}$, however, the solution can still be found by using
convex optimization techniques. Similar equation to Eq.~(\ref{reg})
has been derived in the context of reconstructing a sparse
asymmetric, asynchronous Ising network~\cite{Roudi-12}. Here we
derive the inference equation (Eq.~(\ref{reg})) for the static
reconstruction of a sparse network. We will show in the next section
the efficiency of this hybrid strategy to improve the prediction of
the BA without regularization. To evaluate the efficiency, we define
the reconstruction error (root mean squared (rms) error) as
\begin{equation}\label{error}
\Delta_{J}=\left[\frac{2}{N(N-1)}\sum_{i<j}(J_{ij}^{*}-J_{ij}^{{\rm
true}})^{2}\right]^{1/2}
\end{equation}
where $J_{ij}^{*}$ is the inferred coupling while $J_{ij}^{{\rm
true}}$ is the true one constructed according to Eq.~(\ref{J_spar}).
Other performance measures for sparse network inference will also be
discussed in the following section.
\section{Results and discussions}
\label{sec_result}

\begin{figure}
    \includegraphics[bb=23 12 296 219,width=8.5cm]{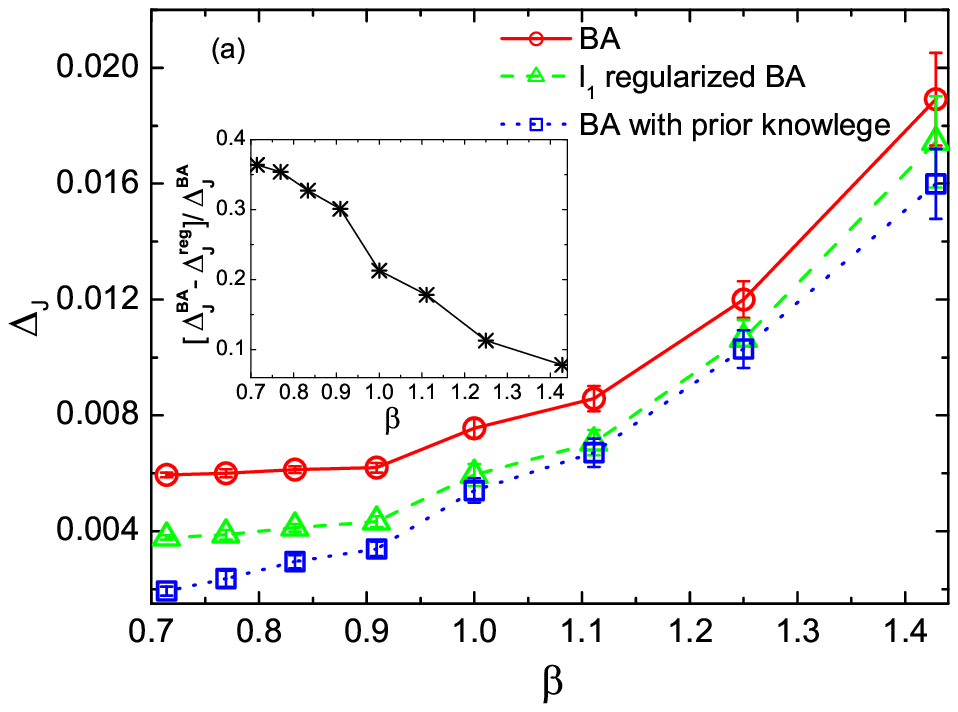}\vskip .1cm
    \includegraphics[bb=16 17 280 212,width=8.5cm]{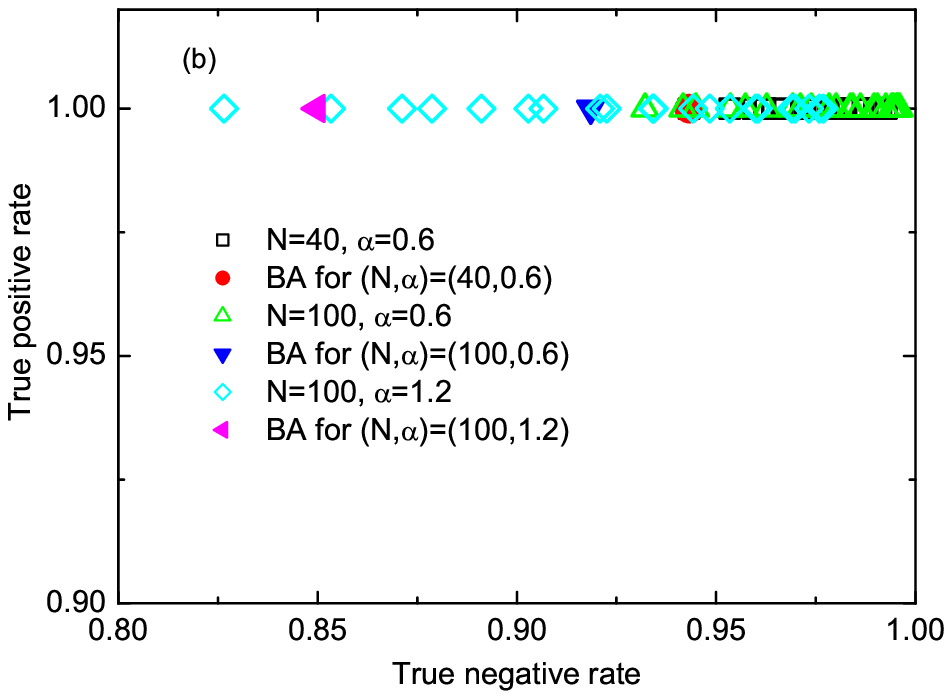}\vskip .1cm
  \caption{
  (Color online) (a) Improvement of the prediction by $\ell_{1}$-regularized BA on sparse Hopfield
  networks. The inference error by BA with prior knowledge of the sparseness of the network is also shown. Network size $N=100$, the memory load $\alpha=0.6$ and
  the mean node degree $l=5$. Each data point is the average over
  five random sparse networks. The regularization parameter has been
  optimized. The inset gives the relative inference error defined as $\frac{\Delta_{J}^{BA}-\Delta_{J}^{reg}}{\Delta_{J}^{BA}}$ versus the inverse
  temperature. (b) The receiver operating characteristic curve for
  three
  typical examples ($T=1.4$). Each data point corresponds to a value of
  $\lambda$ for $\ell_{1}$-regularized BA. The solid symbol gives the result of BA without
  regularization. Parameters for these three examples are
  $(N,P,\alpha)=(40,3,0.6),(100,3,0.6),(100,5,1.2)$ respectively.
  }\label{regBA}
\end{figure}

\begin{figure}
    \includegraphics[bb=18 18 303 220,width=8.5cm]{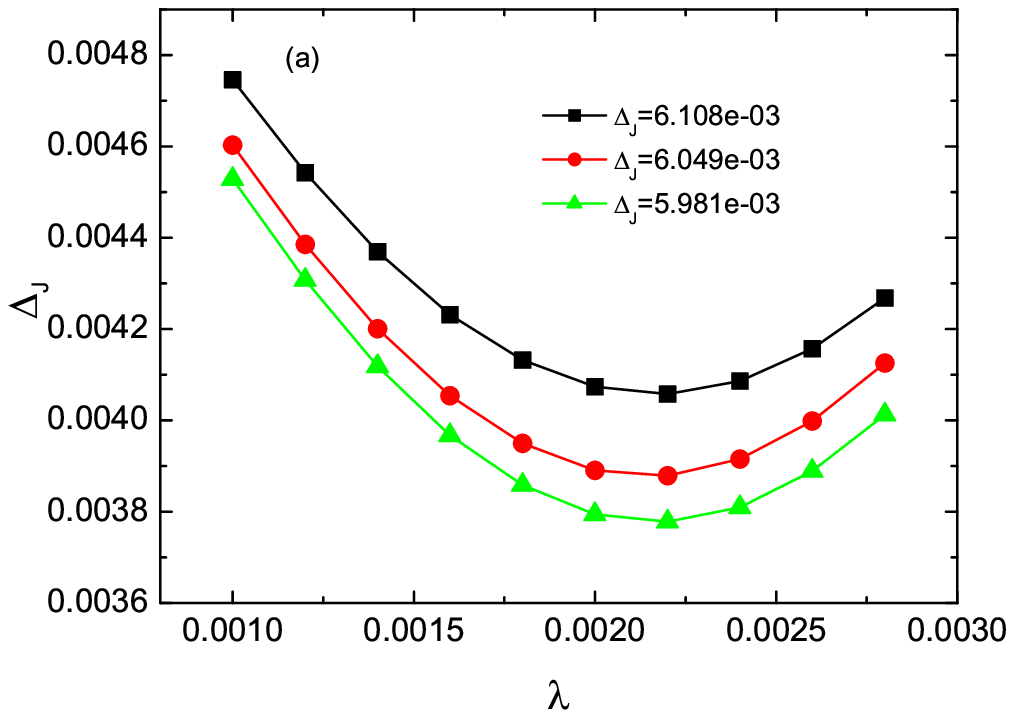} \vskip .1cm
    \includegraphics[bb=18 16 294 214,width=8.5cm]{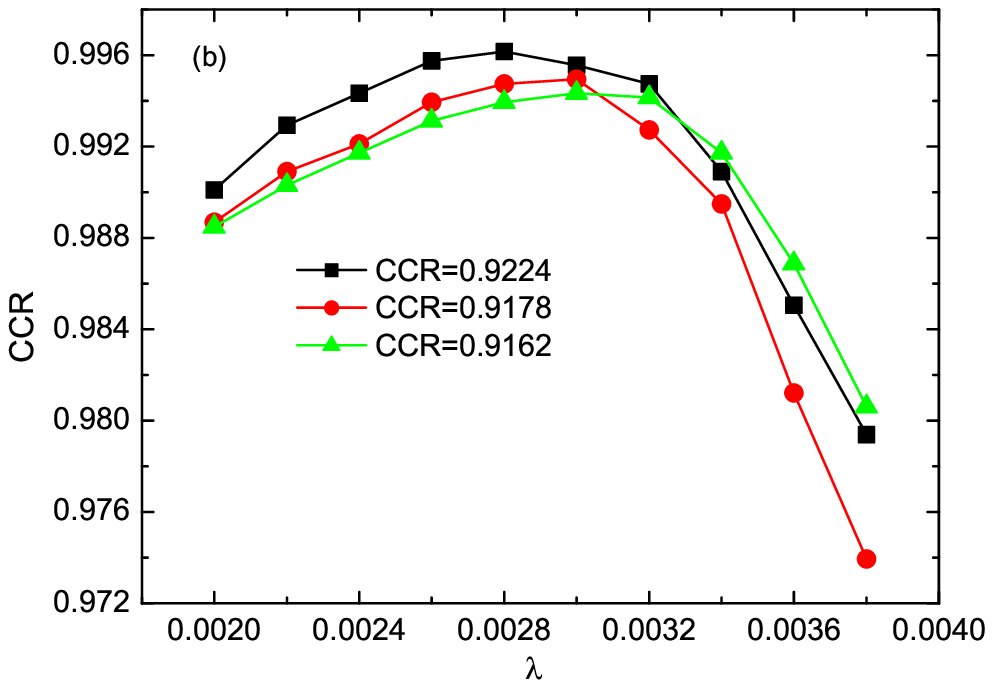}\vskip .1cm
    \includegraphics[bb=16 11 297 216,width=8.5cm]{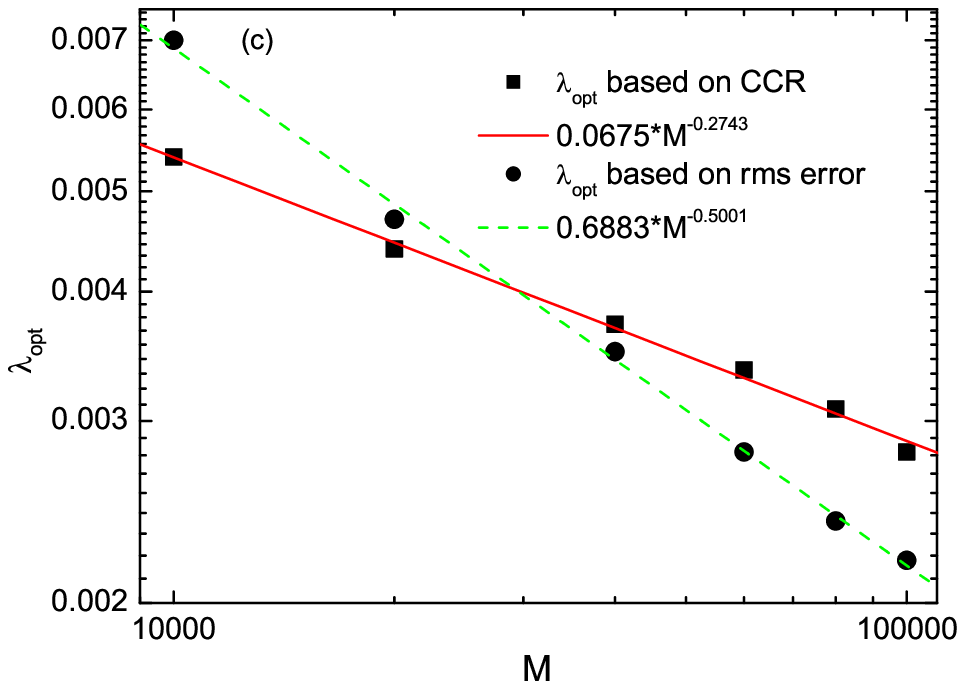}\vskip .1cm
  \caption{
  (Color online) (a) Reconstruction error $\Delta_{J}$ versus the regularization
  parameter $\lambda$ at $T=1.4$. Inference results on three random instances are
  shown. The inference errors by applying BA without regularization on these three random
  instances are $\Delta_{J}=0.006108,0.006049,0.005981$
  respectively. (b) Correct classification rate (CCR) versus the regularization parameter $\lambda$ at $T=1.4$. The instances are the same as those in
  (a). The CCR of BA without regularization are ${\rm CCR}=0.9224,0.9178,0.9162$ respectively.  (c) The optimal $\lambda_{{\rm opt}}$ versus the
  number of samples $M$ ($T=1.4$). Each point is the mean value over five
  random realizations of the sparse Hopfield network. The standard
  error is nearly zero and not shown. The linear fit shows that $\lambda_{{\rm
  opt}}=\lambda_{0}M^{-\nu}$ with
  $\lambda_{0}\simeq0.6883,\nu\simeq0.5001$ for rms error and $\lambda_{0}\simeq0.0675,\nu\simeq0.2743$ for CCR measure.
  }\label{regpara}
\end{figure}
We simulate the sparsely connected Hopfield network of size $N=100$
at different temperatures. The average connectivity for each neuron
$l=5$ and the memory load $\alpha=0.6$. As shown in fig.~\ref{regBA}
(a), the $\ell_{1}$-regularization in Eq.~(\ref{reg}) does improve
the prediction on the sparse network reconstruction. The improvement
is evident in the presence of high quality data (e.g., in the high
temperature region, see  the inset of fig.~\ref{regBA} (a)).
However, the relative inference error (improvement fraction) shown
in the inset of fig.~\ref{regBA} (a) gets smaller as the temperature
decreases. This may be due to insufficient
samplings~\cite{Huang-2012} of glassy states at the low
temperatures. The glassy phase is typically characterized by a
complex energy landscape exhibiting numerous local minima. As a
result, the phase space we sample develops higher order (higher than
second order) correlations whose contributions to the regularized
cost function can not be simply neglected, which explains the
behavior observed in the inset of fig.~\ref{regBA} (a). In this
case, the pseudo-likelihood method or more complex selective cluster
expansion can be used at the expense of larger computation times.
For comparison, we also show the inference error of BA with prior
knowledge of the network connectivity, i.e., the sparseness is known
in advance with only the true non-zero couplings to be predicted.
The comparison confirms that the $\mathbf{C}^{i}$ matrix obtained
from correlations in the data contains useful information about the
sparsity of the network, and this information can be extracted by
using $\ell_{1}$-regularization in Eq.~(\ref{reg}).

An accurate pruning of the network can be achieved by simple
thresholding (setting to zero some couplings whose absolute values
are below certain threshold) based on the improved prediction. The
receiver operating characteristic (ROC) curves are given in
fig.~\ref{regBA} (b) for three typical examples of different network
size, memory load and connectivity. The ROC curve is obtained by
plotting true positive rate (the number of inferred non-zero
couplings with correct sign divided by the total number of true
non-zero couplings) against true negative rate (the number of
inferred zero couplings divided by the total number of true zero
couplings). A threshold $\delta=0.01$ is used to get the inferred
zero couplings. The ROC curve in fig.~\ref{regBA} (b) shows that one
can push the inference accuracy towards the upper right corner (high
true positive rate as well as high true negative rate) by tuning the
regularization parameter. Note that BA without regularization
reports low true negative rate.

We also explore the effects of the regularization parameter on the
reconstruction, which are reported in fig.~\ref{regpara} (a). With
increasing $\lambda$, the inference error first decreases, then
reaches a minimal value followed by an increasing trend in the range
we plot in fig.~\ref{regpara} (a). This implies that the optimal
regularization parameter guides our inference procedure to a sparse
network closest to the original one. The inference quality can also
be measured by the fraction of edges $(ij)$ where the coupling
strength is classified correctly as \textquoteleft
positive\textquoteright, \textquoteleft zero\textquoteright  or
\textquoteleft negative\textquoteright. We call this quantity
correct classification rate (CCR). Results for three typical
examples are reported in fig.~\ref{regpara} (b). With increasing
$\lambda$, CCR first increases and then decreases. The optimal
regularization parameter corresponding to the maximum is slightly
different from that in fig.~\ref{regpara} (a). By using regularized
BA (Eq.~(\ref{reg})), one can achieve a much higher value of CCR,
and furthermore the computational cost is not heavy. Interestingly,
the optimal value of $\lambda$ yielding the lowest inference error
(rms error) has the order of $\mathcal {O}(\sqrt{\frac{1}{M}})$ for
fixed network size (usually $M\gg N$), which is consistent with that
found in Refs.~\cite{Wain-2010,Bento-2011}. We verify this scaling
form by varying $M$ and plotting the optimal $\lambda$ in
fig.~\ref{regpara} (c). The linear fit implies that the scaling
exponent $\nu\simeq0.5$. However, this scaling exponent depends on
the performance measure. Taking the CCR measure yields a smaller
value $\nu\simeq0.2743$, as shown in fig.~\ref{regpara} (c) as well.
We also find that the magnitude of the optimal regularization
parameter shows less sensitivity to specific instances and other
parameters (e.g., the temperature, memory load or network size),
since the number of samplings $M$ dominates the order of the
magnitude. The specific optimal value becomes slightly different
across different instances of the sparse network in the low
temperature region, where its mean value shifts to a bit larger
value for rms error measure or a bit smaller value for CCR measure,
as the temperature further decreases. The number of samplings $M$
determines the order of the magnitude, which helps us find the
appropriate strength for the regularization parameter. In the real
application, the true coupling vector is {\it a priori} unknown. In
this case, the regularization parameter can be chosen to make the
difference between the measured moments and those produced by the
reconstructed Ising model as small as possible.

\begin{figure}
    \includegraphics[bb=16 12 285 216,width=8.5cm]{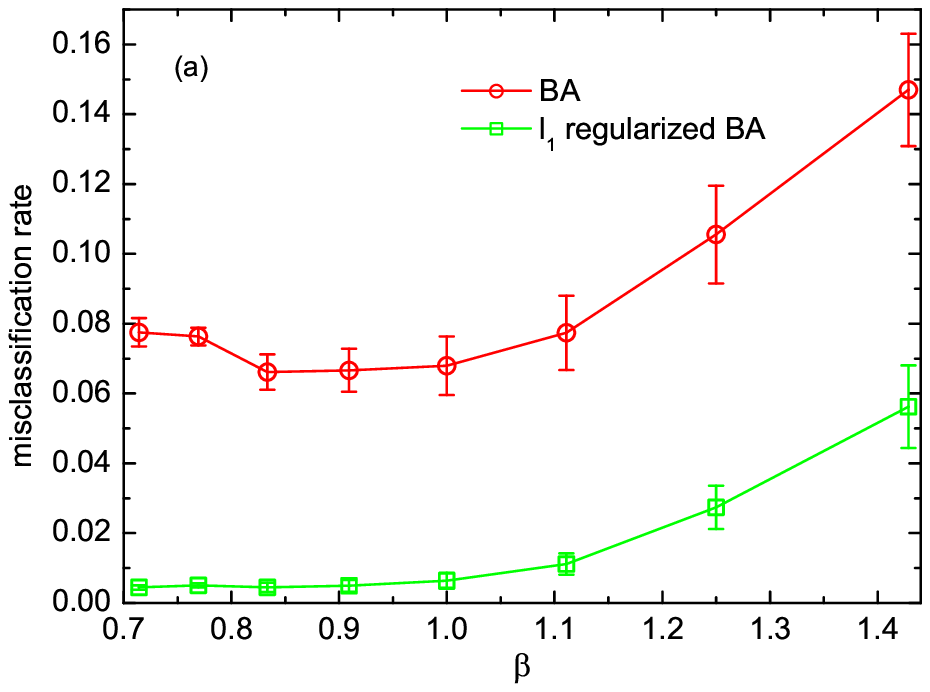}\vskip .1cm
    \includegraphics[bb=18 17 285 216,width=8.5cm]{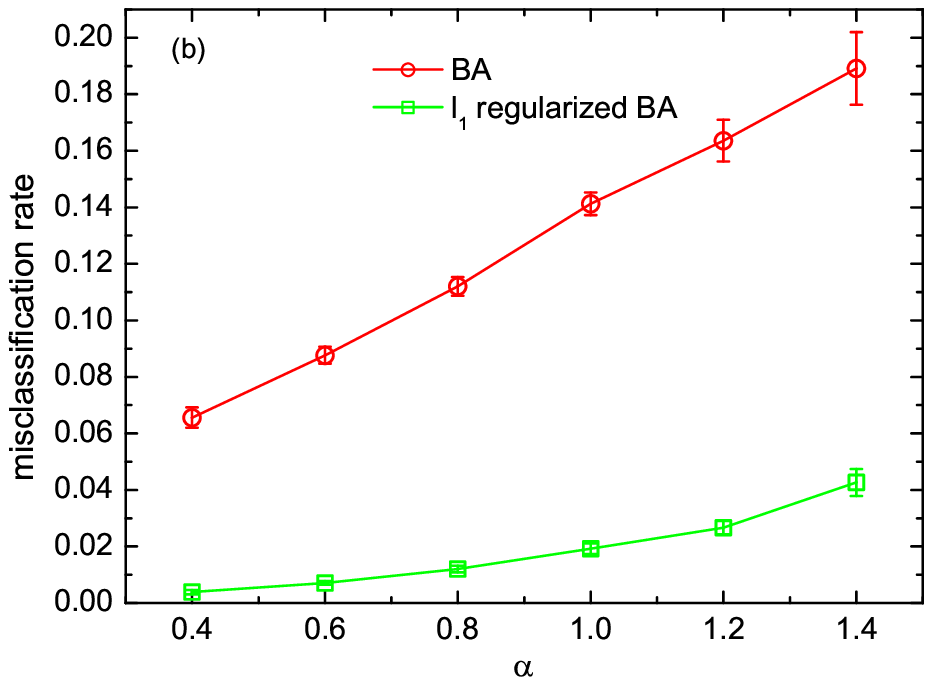}\vskip .1cm
  \caption{
  (Color online) Comparison of performance measured by misclassification rate. Each data point is the average over
  five random sparse networks. The regularization parameter has been
  optimized. (a) Misclassification rate versus inverse temperature. Network size $N=100$, the memory load $\alpha=0.6$ and
  the mean node degree $l=5$. (b) Misclassification rate versus memory load. Network size $N=100$, temperature $T=1.4$ and $P=5$.
  }\label{MisCR}
\end{figure}

Finally, we give the comparison of performance measured by
misclassification rate in fig.~\ref{MisCR}. According to the above
definition, misclassification rate equals to $1-{\rm CCR}$. Low
misclassification rate is preferred in the sparse network inference.
Fig.~\ref{MisCR} (a) shows the performance versus inverse
temperature. The misclassification rate is lowered by a substantial
amount using the hybrid strategy. Especially in the high temperature
region, the error approaches zero while BA still yields an error of
the order of $\mathcal {O}(10^{-2})$. As displayed in
fig.~\ref{MisCR} (b), the hybrid strategy is also superior to BA
when the memory load is varied, although the misclassification rate
grows with the memory load. Compared with BA, the
$\ell_{1}$-regularized BA yields a much slower growth of the error
when $\alpha$ increases. Even at the high memory load $\alpha=1.4$,
the hybrid strategy is able to reconstruct the network with an error
$4.3\%$ while at the same memory load, the error of BA is as large
as $18.9\%$. Note that as $\alpha$ changes, the average connectivity
also changes. Fig.~\ref{MisCR} (b) illustrates that our simple
inference strategy is also robust to different mean node degrees.

\section{Conclusion}
\label{sec_Sum}

We propose an efficient hybrid inference strategy for reconstructing
the sparse Hopfield network. This strategy combines Bethe
approximation and the $\ell_{1}$-regularization by expanding the
objective function (negative log-likelihood function) up to the
second order of the coupling fluctuation around its (near) optimal
value. The hybrid strategy is simple in implementation without heavy
computational cost, yet improves the prediction by zeroing couplings
which are actually not present in the network (see fig.~\ref{regBA}
and fig.~\ref{MisCR}). We can control the accuracy by tuning the
regularization parameters. The magnitude of the optimal
regularization parameters is determined by the number of independent
samples $M$ as $\lambda_{{\rm opt}}\sim M^{-\nu}$.  The value of the
scaling exponent depends on the performance measure. $\nu\simeq0.5$
for root mean squared error measure while $\nu\simeq0.2743$ for
misclassification rate measure. By varying the value of the
regularization parameter, we show that the reconstruction (rms)
error first decreases and then increases after the lowest error is
reached. Similar phenomenon is observed for the change of
misclassification rate with the regularization parameter. We observe
this phenomenon in the sparse Hopfield network reconstruction, and
this behavior may be different in other cases~\cite{Cocco-12}. The
efficiency of this strategy is demonstrated for the sparse Hopfield
model, but this approximated reconstruction method is generally
applicable to other diluted mean field models if we can first find a
good solution (yielding low inference error) to the inverse Ising
problem without regularization.


\section*{Acknowledgments}

Helpful discussions with Yoshiyuki Kabashima are gratefully acknowledged. This work was supported by the JSPS Fellowship for Foreign
Researchers (Grant No. $24\cdot02049$).



\end{document}